\documentclass[a4paper]{article}
\usepackage[latin9]{inputenc}
\usepackage{verbatim}
\usepackage{amsmath}
\usepackage{graphicx}

\makeatletter

\pdfpageheight\paperheight
\pdfpagewidth\paperwidth

\providecommand{\tabularnewline}{\\}

\usepackage{INTERSPEECH2022}
\usepackage[bookmarks=false,pdftex,pdfpagemode=None,pdfstartview=FitH,pdfview=FitH,pdftitle={Real-Time Packet Loss Concealment With Mixed Generative and Predictive Model},pdfauthor={Jean-Marc Valin, Ahmed Mustafa, Christopher Montgomery, Timothy B. Terriberry, Michael Klingbeil, Paris Smaragdis, Arvindh Krishnaswamy}]{hyperref}
\usepackage{balance}

\makeatother

\begin{document}
\name{Jean-Marc Valin$^\natural$, Ahmed Mustafa$^\natural$, Christopher Montgomery$^\natural$, Timothy B. Terriberry$^\natural$,\\Michael Klingbeil$^\natural$, Paris Smaragdis$^\natural$$^\flat$, Arvindh Krishnaswamy$^\natural$}

\address{$^\natural$Amazon Web Services\quad$^\flat$University of Illinois at Urbana-Champaign}

\email{\{jmvalin, ahdmust, chrimonh, territim, klingm, parsmara, arvindhk\}@amazon.com}
\title{Real-Time Packet Loss Concealment With Mixed Generative and Predictive
Model}

\maketitle
\maketitle 
\begin{abstract}
As deep speech enhancement algorithms have recently demonstrated capabilities
greatly surpassing their traditional counterparts for suppressing
noise, reverberation and echo, attention is turning to the problem
of packet loss concealment (PLC). PLC is a challenging task because
it not only involves real-time speech synthesis, but also frequent
transitions between the received audio and the synthesized concealment.
We propose a hybrid neural PLC architecture where the missing speech
is synthesized using a generative model conditioned using a predictive
model. The resulting algorithm achieves natural concealment that surpasses
the quality of existing conventional PLC algorithms and ranked second
in the Interspeech 2022 PLC Challenge. We show that our solution not
only works for uncompressed audio, but is also applicable to a modern
speech codec.
\end{abstract}

\section{Introduction}

Real-time voice communication over the Internet typically relies on
``best-effort'' unreliable transmission protocols (RTP/UDP) to minimize
latency. When voice packets are lost (or arrive too late to be played
back), the receiver attempts to\emph{ conceal} the loss in such a
way as to limit the quality degradation caused by the loss. Traditional
packet loss concealment (PLC) techniques often involve repeating pitch
periods~\cite{sanneck1996new}, which improves quality over filling
the missing packets with silence, but often also causes noticeable
artifacts. 

Recent techniques based on deep neural networks (DNN) have been shown
to significantly improve quality for both noise suppression~\cite{reddy2020interspeech}
and echo cancellation~\cite{aec_challenge}. DNN techniques for neural
PLC have also recently been investigated~\cite{lee2015packet,Lotfidereshgi2018SpeechPU,mohamed2020concealnet,stimberg2020waveneteq,lin2021convolutional,pascual2021adversarial}
and in this work, we propose a new PLC algorithm that ranked second
in the Interspeech 2022 Audio Deep Packet Loss Concealment Challenge~\cite{plc_challenge}. 

We propose a hybrid architecture that combines both a generative vocoder
and a predictive conditioning model (Section~\ref{sec:Neural-Concealment-Architecture}).
During losses, the predictive model estimates the acoustic features
to be used by the generative LPCNet neural vocoder (Section~\ref{sec:Conditioning-Prediction}).
Synthesis is performed in the time domain and in Section~\ref{sec:Framing,-Transitions-Resync}
we proposed both causal and non-causal approaches to post-loss resynchronization.
The evaluation in Section~\ref{sec:Evaluation-Results} demonstrates
that the proposed real-time PLC algorithm significantly out-performs
conventional PLC algorithms.

\section{Neural Concealment Architecture}

\label{sec:Neural-Concealment-Architecture}

To conceal missing packets, we want to use some form of neural vocoder,
with many options available, including GAN-based~\cite{donahue2019wavegan},
flow-based~\cite{prenger2019waveglow}, and autoregressive~\cite{van2016wavenet}
techniques. Although not strictly necessary, it is desirable for the
vocoder to be able to synthesize speech that perfectly aligns with
the ground-truth speech before a loss, thereby avoiding the need for
a cross-fade at the beginning of the loss. Also, considering that
the challenge rules target CPU inference, highly-parallel synthesis
is not required, making autoregressive vocoders a good choice. In
this work, we use LPCNet~\cite{valin2019lpcnet}.

Purely autoregressive models have been shown to successfully conceal
packet loss, with various measures to counter the model's natural
tendency to drift and babble over time~\cite{stimberg2020waveneteq}.
That behavior is expected since speech is being synthesized with no
conditioning beyond the start of the packet loss.

In this work, we propose a hybrid approach combining generative and
predictive models. A generative autoregressive model is used to conceal
the missing audio samples, but the acoustic features used for generation
are determined using a predictive model. This allows precise long-term
control of the spectral trajectory being reconstructed, while generating
natural-sounding speech.

This can be viewed as different generation criteria for different
time scales. While we want our algorithm to be ``creative'' in extending
missing segments of a phoneme using plausible-sounding audio, we do
\emph{not} want an algorithm that generates more phonemes or words
that have not been spoken -- no matter how plausible.

\subsection{LPCNet Model}

LPCNet is an autoregressive neural vocoder that improves on WaveRNN~\cite{kalchbrenner2018efficient}
by using linear prediction. LPCNet is divided into a frame rate network
and a sample rate network. The frame rate network, operating at 100~Hz
on acoustic features, outputs conditioning parameters for the sample
rate network that autoregressively generates the 16~kHz speech samples.
In this work, we use the improved version of LPCNet with significantly
reduced complexity~\cite{valin2022lpcnetefficiency}.

LPCNet uses acoustic feature vectors based on 20\nobreakdash-ms overlapping
windows with a 10\nobreakdash-ms interval. Each consists of 18~Bark
frequency cepstral coefficients (BFCC), a pitch period, and a pitch
correlation. Each 10\nobreakdash-ms synthesized speech segment corresponds
to the center of the analysis window.

The original LPCNet model uses two 3x1 convolutional layers in its
frame rate network, configured in such a way as to use two feature
vectors ahead of the frame being synthesized. This extra context improves
the synthesis quality at the cost of 25~ms of added latency. For
PLC, when a packet is lost, we do not have the future features and
trying to predict them would result in more unreliable information
being used for the synthesis. For that reason, we use a model with
causal features. That being said, due to the overlap in the analysis
windows, we still have a 5~ms algorithmic delay, as we synthesize
the middle 10~ms of the 20\nobreakdash-ms analysis window. 

\section{Conditioning Prediction}

\label{sec:Conditioning-Prediction}

\begin{figure}
\begin{centering}
\includegraphics[width=1\columnwidth]{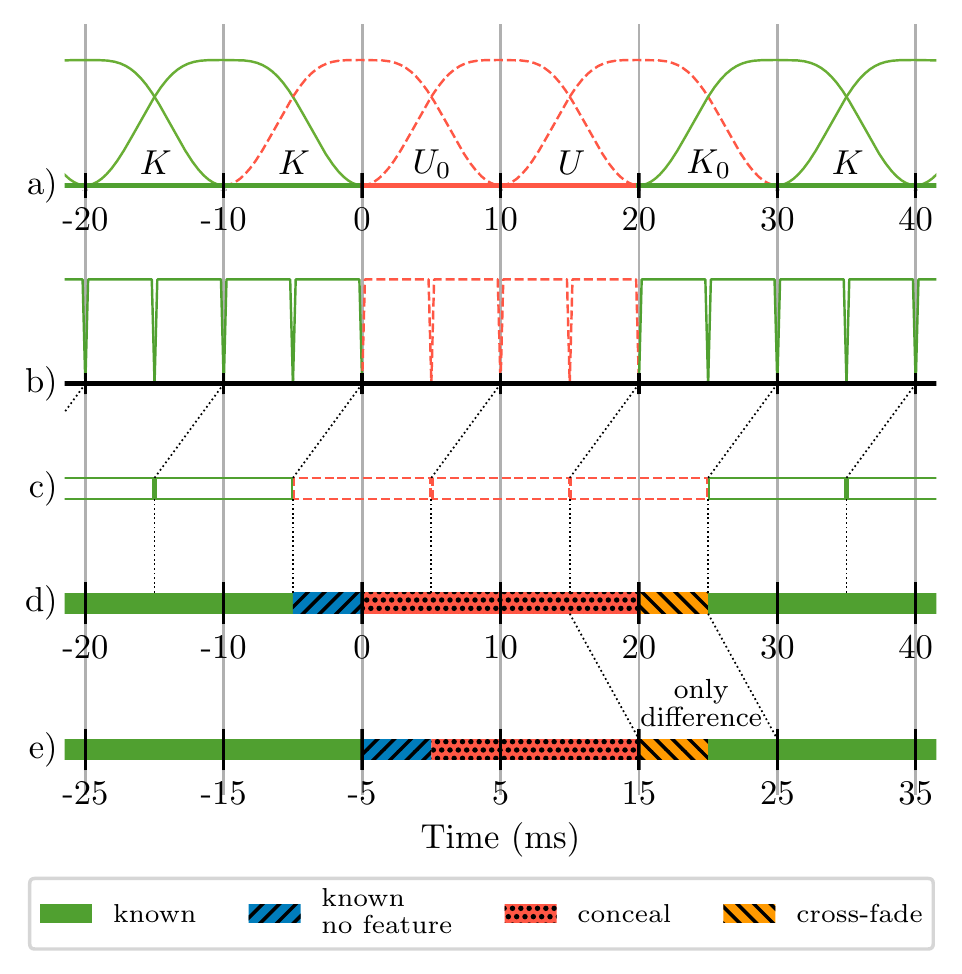}
\par\end{centering}
\caption{Proposed PLC for a sequence where the samples at $t\in\left[0\,\mathrm{ms},20\,\mathrm{ms}\right[$
are unknown (lost): a) feature analysis windows, with the dashed red
windows denoting the features that cannot be computed due to missing
data b) Burg analysis segments (Section~\ref{subsec:Improved-Temporal-Resolution}),
where the dashed red segments cannot be computed c) LPCNet input features,
where the green features are a copy of the features in a) and the
dashed red feature vectors are predicted by the DNN model d) output
samples for causal processing (Section~\ref{subsec:Causal}), where
the green samples are known and used to update the LPCNet state with
known features, the blue samples are known, but update the state using
predicted features, the red samples are synthesized by LPCNet, and
the orange samples are a cross-fade of synthesized and known samples
e) non-causal processing (Section~\ref{subsec:Non-causal}) illustrating
the 5\protect\nobreakdash-ms delay and using the same colors, except
that orange samples are a cross-fade of forward and backward synthesis.
The tick marks still denote the processing intervals. Note the shorter
total concealment needed for the non-causal case. \label{fig:Features-and-framing}}
\end{figure}

LPCNet requires features at a 10\nobreakdash-ms interval to condition
the synthesis. When packets are lost, we cannot compute those features
from the missing data, so we need to estimate them based on the received
samples. The simplest approach is to just \emph{freeze} the features
during loss, i.e. use the last feature vector available. Unfortunately,
when packet loss occurs during non-stationary periods, the concealment
produces noticeable artifacts. For example, when a packet is lost
at the end of a syllable, the last feature vector represents the energy
centered 10~ms before the loss, and repeating it causes an audible
burst in energy. 

Instead, we use a DNN model to predict the unknown feature vectors
from the known ones. To provide as much context for the prediction
as possible, we use a recurrent neural network (RNN). For each frame,
the input is either the known feature vector, or a vector of zeros.
We include a ``lost'' binary flag to the input vector since the
known features could theoretically be all zeros. The prediction network
consists of one fully-connected layer, followed by two GRUs~\cite{cho2014properties},
and a fully-connected output layer.

\subsection{Perceptual Loss Functions}

\label{subsec:Perceptual-Loss-Functions}

Because some aspects of the PLC task are completely unpredictable,
we need a robust loss that treats unexpected/unpredictable events
like label noise. For that reason, an $L_{1}$ loss function is more
appropriate than an $L_{2}$ loss. We find that we can further improve
quality by using a different loss for the different predicted features. 

For the cepstral features, we find that in addition to an $L_{1}$
loss in the cepstral domain we can benefit from considering the bands
themselves. Moreover, we find that overestimating the energy of voiced
frames hurts quality more than underestimating it, so we adaptively
bias the spectral loss:
\begin{equation}
\mathcal{L}_{s}=\mathbb{E}\!\left[\sum_{k}\left(\left|\Delta c_{k}\right|\!+\!\left|\Delta b_{k}\right|\!+\!\alpha\max\!\left(\Delta b_{k},0\right)\!\right)\right]\,,\label{eq:spectral_loss}
\end{equation}
where $\Delta c_{k}=\hat{c}_{k}-c_{k}$ is the difference between
the predicted and ground truth cepstral coefficients, $\Delta b_{k}$
is the difference in band energies computed from the inverse discrete
cosine transform (IDCT) of $\Delta c_{k}$, and $\alpha$ is a bias
coefficient equal to~1 for voiced frames and~0 for unvoiced frames.

Because the pitch period is always included in the features (even
for unvoiced frames), it is particularly noisy and its loss function
must be particularly robust to outliers. We use 
\begin{equation}
\mathcal{L}_{p}=\mathbb{E}\left[\left|\Delta p\right|+\beta_{1}\min\left(\left|\Delta p\right|,50\right)+\beta_{2}\min\left(\left|\Delta p\right|,20\right)\right]\ ,\label{eq:pitch_loss}
\end{equation}
where $\Delta p=\hat{p}-p$ is the difference between the predicted
and ground truth pitch period and $\beta_{\times}$ are tuning parameters
(we use $\beta_{1}=20$ and $\beta_{2}=160$).

We find that overestimating the pitch correlation feature $r$ leads
to better pitch stability so we use the biased loss
\begin{equation}
\mathcal{L}_{c}=\mathbb{E}\left[\left|\Delta_{r}\right|+2\max\left(-\Delta_{r},0\right)\right]\ .\label{eq:corr_loss}
\end{equation}

\subsection{Improved Temporal Resolution}

\label{subsec:Improved-Temporal-Resolution}

One limitation with the LPCNet features in the context of PLC is that
the 20\nobreakdash-ms analysis window is centered 10~ms before the
concealment begins (Fig.~\ref{fig:Features-and-framing}a), so it
does not always capture the changes that occur just before a loss.
To address the problem, we introduce additional features based on
short-term spectral analysis. Burg spectral estimation~\cite{burg1975}
is well-suited for such analysis since it does not require windowing.
We estimate the Burg spectrum independently on each 5\nobreakdash-ms
half of each new frame, as shown in Fig.~\ref{fig:Features-and-framing}b.
The Burg-derived all-pole filters are then converted into cepstral
coefficients that supplement the input to the feature prediction DNN
(Fig.~\ref{fig:Features-and-framing}c). We can also see from Fig.~\ref{fig:Features-and-framing}
that the loss of $k$ frames results in the loss of $\left(k+1\right)$
LPCNet feature vectors, but only $k$ Burg feature vectors, which
is another benefit of the Burg analysis proposed here.

\subsection{Handling Long Bursts}

During long bursts of lost packets, it is meaningless to attempt concealment
beyond around 100~ms. In that case, it is desirable to fade the concealment
in the least objectionable way. Fading too slowly results in artifacts
that sound similar to heavy breathing, whereas fading too quickly
sounds very unnatural. We choose to fade out in such a way as to match
the reverberation time of a small room with $\mathrm{RT}_{60}=120\,\mathrm{ms}$.
This can be accomplished, at inference time, by linearly decreasing
the first predicted cepstral coefficient ($c_{0}$) over time after
the initial 100~ms. As a result, long losses sound similar to a talker
being naturally interrupted.

\section{Framing, Transitions \& Resynchronization}

\label{sec:Framing,-Transitions-Resync}

Unlike many other applications of neural vocoders, the PLC task involves
using both known and synthesized audio in its output. It is thus important
to make the transitions as seamless as possible. We consider four
types of frames: known frames ($K$), transition unknown/lost frames
at the beginning of a loss ($U_{0}$), unknown frames ($U$), and
transition known frames after a loss ($K_{0}$). Each frame type requires
different processing. Also, in this work, we propose three ways of
performing the concealment: causal, non-causal, and inside a stateful
codec.

\subsection{Causal}

\label{subsec:Causal}

The causal concealment case, with no look-ahead allowed, is depicted
in Fig.~\ref{fig:Features-and-framing}a\nobreakdash-\ref{fig:Features-and-framing}d.
When no loss is occurring, the incoming $K$ frame containing samples
$\left[t-10\,\mathrm{ms},t\right]$ is used to compute both the LPCNet
features from the window covering $\left[t-20\,\mathrm{ms},t\right]$,
as well as the two Burg features in $\left[t-10\,\mathrm{ms},t\right]$.
The features are used to update the prediction DNN state (the output
is discarded). The LPCNet features are also used alongside samples
$\left[t-15\,\mathrm{ms},t-5\,\mathrm{ms}\right]$ to update the state
of the LPCNet vocoder, meaning that the samples are given to the network
as input, but the output probability is discarded. 

Upon the first loss, we need to predict two LPCNet feature vectors.
The first is used to continue updating the LPCNet state using known
samples up to time $t$ and then to synthesize the $U_{0}$ samples
up to $t+5\,\mathrm{ms}$. The second predicted vector is used to
synthesize samples $\left[t+5\,\mathrm{ms},t+10\,\mathrm{ms}\right]$.
For any further loss only one new vector is predicted and the first
half of the $U$ samples are synthesized using the ``old'' feature
vector and the second half is synthesized using the new vector. 

On the first new packet after a loss, the samples at $\left[t+20\,\mathrm{ms},t+30\,\mathrm{ms}\right]$
(in our example) cannot be used to compute LPCNet features (since
the first half of the window is lost), but can be used for Burg features.
In turn, the Burg features can be used to update the latest prediction
from the DNN. That prediction is used to continue the synthesis up
to $t+25\,\mathrm{ms}$, cross-fading these generated samples with
the $K_{0}$ samples to avoid any discontinuity. 

\subsection{Non-causal}

\label{subsec:Non-causal}

In the non-causal PLC case (Fig.~\ref{fig:Features-and-framing}e),
we improve the concealment quality by using 5~ms of look-ahead in
the processing. When packets are continuously received, the processing
is the same as the causal case, except that the algorithm delays the
output signal by 5~ms to keep its look-ahead buffer. Similarly, there
are no fundamental changes during loss. The improvement comes from
the resynchronization after a loss. When the first packet arrives
after a loss, we use the 10~ms of speech in the packet to extrapolate
the speech backwards in time for 5~ms. Because of the delay, that
5~ms segment has not yet been played, so we can cross-fade the backward
and forward signal extensions. By doing so, we are able to play the
all received audio packets unmodified, unlike in the causal case.

\subsection{Stateful codec}

Both the causal and non-causal PLC cases apply to situations where
audio is either uncompressed, or compressed with a simple stateless
codec such as G.711~\cite{G711}. When speech is compressed using
a stateful codec such as Opus~\cite{rfc6716} or AMR-WB~\cite{bessette2002adaptive},
the concealed audio is needed to reconstruct the first packet after
a loss, so the non-causal concealment proposed above is not possible.
Moreover, because the linear prediction used in most modern codecs
inherently avoids discontinuities, the cross-fade step from the causal
case is not needed. Aside from that, the rest of Fig.~\ref{fig:Features-and-framing}d
also holds for the stateful codec case. In this work, we demonstrate
a neural PLC operating inside the voice coding mode of Opus~\cite{vos2013}.
In that context, our proposed PLC algorithm completely replaces the
existing SILK PLC and its output is also used for prediction (long-
and short-term) in the decoder state when a new packet arrives.

\section{Evaluation \& Results}

\label{sec:Evaluation-Results}

\begin{table}
\caption{Computation time (in ms) to process 10~ms of speech for each type
of frame, as measured on an Intel i7-10810U laptop CPU.\label{tab:Computation-time}}

\begin{centering}
\begin{tabular}{lcccc}
\hline 
Algorithm & $K$~~ & $U_{0}$ & $U$ & $K_{0}$\tabularnewline
\hline 
Causal & 1.35~~ & 2.12 & 1.34 & 1.53\tabularnewline
Non-Causal & 1.38{*} & 1.33 & 1.34 & 2.58\tabularnewline
Codec & 1.38~~ & 2.18 & 1.37 & 0.84\tabularnewline
\hline 
\end{tabular}
\par\end{centering}
\centering{}{*}2.54~ms for the first frame following a $K_{0}$ frame
\end{table}

The LPCNet vocoder model is trained on 205~hours of 16\nobreakdash-kHz
speech from a combination of TTS datasets~\cite{demirsahin-etal-2020-open,kjartansson-etal-2020-open,kjartansson-etal-tts-sltu2018,guevara-rukoz-etal-2020-crowdsourcing,he-etal-2020-open,oo-etal-2020-burmese,van-niekerk-etal-2017,gutkin-et-al-yoruba2020,bakhturina2021hi}
including more than 900~speakers in 34~languages and dialects. The
training is performed as described in~\cite{valin2022lpcnetefficiency},
except that we explicitly randomize the sign of each training sequence
so that the algorithm works for any polarity of the speech signal.
Due to recent improvements to the LPCNet efficiency, we are able to
use a $\mathrm{GRU_{A}}$ size of 640~units, with 15\% density, while
meeting all real-time constraints. 

The feature prediction model is trained on all 205~hours used for
the vocoder, plus 64~hours of training speech published by the PLC
challenge organizers. The size of the GRUs is set to 512~units, with
a 256-unit input fully-connected layer. The source code for the proposed
PLC is available under an open-source license at \url{https://github.com/xiph/LPCNet}
in the \texttt{plc\_challenge} branch, with the corresponding Opus
neural PLC at \url{https://gitlab.xiph.org/xiph/opus} in the \texttt{neural\_plc}
branch.

Since the PLC challenge allows for up to 20~ms total latency and
does not involve a codec, we submitted the non-causal PLC for consideration.
Despite that, we still evaluate all use cases in this section. 

\subsection{Complexity}

The complexity of the algorithm is dominated by the complexity of
the LPCNet vocoder, with the feature prediction contributing less
than 20\% of the total complexity. The processing time for each 10\nobreakdash-ms
frame varies over time according to the number of samples processed
by the vocoder based on the loss pattern and the use-case considered
in Section~\ref{sec:Framing,-Transitions-Resync}. The measured processing
times are shown in Table~\ref{tab:Computation-time}. In steady-state
operation (either $K$ or $U$), the algorithm requires between 1.3
and 1.4~ms to process 10~ms of speech, which amounts to between
13\% and 14\% of one CPU core required for real-time operation. The
worst-case computation time for the non-causal case is 2.58~ms. When
considering the 10\nobreakdash-ms frame size and the 5\nobreakdash-ms
look-ahead, the total delay of the non-causal PLC is 17.58~ms, meeting
the 20\nobreakdash-ms maximum allowed for the challenge.

\subsection{Quality}

The PLC challenge organizers~\cite{plc_challenge} evaluated \emph{blind}
test utterances processed with the non-causal PLC algorithm. The comparison
category rating~(CCR) mean opinion score (CMOS)~\cite{P.800} results
were obtained using the crowdsourcing methodology described in P.808~\cite{P.808,naderi2020open},
where 5 randomly-selected listeners were asked to evaluate each of
the 966~test utterances. The 95\% confidence intervals on the CMOS
values are around 0.035. Among the 7~submissions to the challenge,
the proposed algorithm ranked second in the CMOS evaluation and first
on the speech recognition word accuracy evaluation (over the same
test set), taking the second place overall. 

\begin{table}

\caption{PLC Challenge official results. The overall score only considers the
subjective quality evaluation (CMOS) and the speech recognition word
accuracy (WAcc). The proposed algorithm ranks first for word accuracy
and second for both CMOS and overall score. We provide the scores
for the first overall submission and the average of the two submissions
tied for third place. The PLCMOS~\cite{plc_challenge} and DNSMOS~\cite{reddy2022dnsmos}
objective metrics are included only for reference and are not considered
in the rankings.\label{tab:PLC-Challenge-official-results}}

\begin{centering}
\begin{tabular}{lccccc}
\hline 
Algorithm & PLC- & DNS- & \textbf{CMOS} & \textbf{WAcc} & \textbf{Score}\tabularnewline
 & MOS & MOS &  &  & \tabularnewline
\hline 
$\mathrm{1^{st}}$ place & 4.282 & 3.797 & \textbf{-0.552} & 0.875 & \textbf{0.845}\tabularnewline
\textbf{proposed} & 3.744 & 3.788 & -0.638 & \textbf{0.882} & 0.835\tabularnewline
$\mathrm{3^{rd}}$ avg. & 3.903 & 3.686 & -0.825 & 0.864 & 0.794\tabularnewline
Zero-fill & 2.904 & 3.444 & -1.231 & 0.861 & 0.725\tabularnewline
\hline 
\end{tabular}
\par\end{centering}
\end{table}

\begin{table}
\caption{Ablation study results. Compared to feature repetition, we show the
improvement from adding a feature prediction DNN, adding perceptual
loss functions (Section~\ref{subsec:Perceptual-Loss-Functions}),
adding Burg spectral features (Section~\ref{subsec:Improved-Temporal-Resolution}),
and adding 5~ms non-causal processing (Section~\ref{subsec:Non-causal}).\label{tab:Ablation-study}}

\centering{}%
\begin{tabular}{lcc}
\hline 
Algorithm & PESQ-WB & PLC-MOS\tabularnewline
\hline 
Zero-fill & 2.185 & 2.874\tabularnewline
Baseline & 2.059 & 2.786\tabularnewline
\hline 
Repetition & 2.517 & 3.642\tabularnewline
~+DNN (causal) & 2.647 & 3.688\tabularnewline
~~+loss \eqref{eq:spectral_loss}-\eqref{eq:corr_loss} & 2.652 & 3.660\tabularnewline
~~~+Burg & 2.705 & 3.739\tabularnewline
~~~~+\textbf{non-causal} & \textbf{2.766} & 3.790\tabularnewline
\hline 
\end{tabular}
\end{table}

To assess the contribution of different components of the algorithm,
we perform an ablation study using the PESQ-WB~\cite{P.862.2} and
PLCMOS~\cite{plc_challenge} objective metrics. Although PESQ-WB
has been shown in the past not to be suitable for vocoder evaluation,
we have found that for neural PLC, it correlates sufficiently well
with subjective evaluation. Table~\ref{tab:Ablation-study} shows
an ablation study demonstrating the main improvements described in
this work. We note that both metrics agree on all improvements, except
the \eqref{eq:spectral_loss}-\eqref{eq:corr_loss} losses, but we
have found through informal listening that the losses are indeed useful
for improving quality.

In addition to the official challenge experiments, we conducted further
experiments on the challenge development set (because the clean samples
were needed). Those experiments used the same models used for our
challenge submission. In that experiment, we evaluate the causal,
non-causal, and Opus codec cases. To evaluate the Opus PLC, we encoded
the clean utterances at 24~kb/s using the speech mode with 20\nobreakdash-ms
frames, with the encoder robustness settings optimized for 20\% loss
(see Section~3.8 of \cite{vos2013}). We compare the proposed Opus
neural PLC to the default classical PLC. Similarly, we include the
NetEQ\footnote{Based on the WebRTC implementation at \href{https://webrtc.org/}{https://webrtc.org/}}
classical PLC algorithm operating on (uncompressed) PCM audio. 

\begin{figure}
\begin{centering}
\includegraphics[width=0.9\columnwidth]{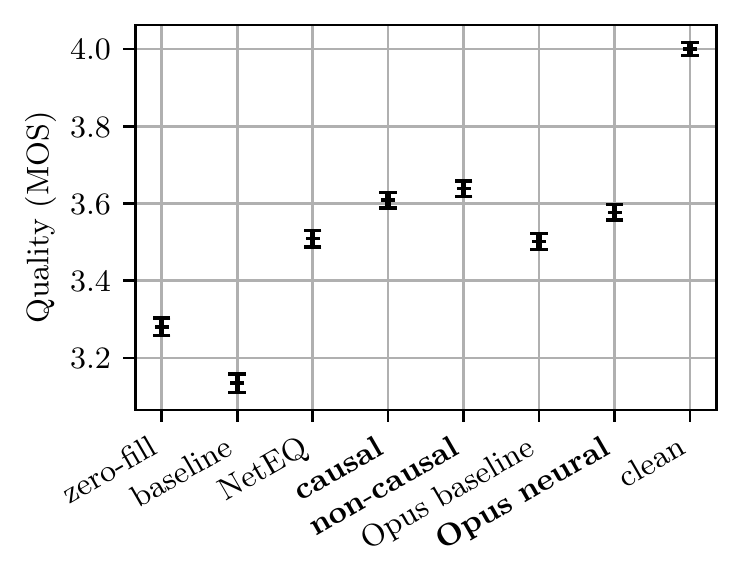}
\par\end{centering}
\caption{Internal MOS results, including the 95\% confidence intervals.\label{fig:Internal-MOS-results}}

\end{figure}

Each of the 966~development utterances was evaluated by 15~randomly-selected
listeners on a MOS absolute category rating~(ACR) scale. The results
in Fig.~\ref{fig:Internal-MOS-results} show that for PCM, the causal
and non-causal versions both out-perform the popular NetEQ PLC and
the baseline PLC. While the difference between the non-causal and
causal algorithm is not statistically significant, the objective results
in Table~\ref{tab:Ablation-study} suggest that non-causal concealment
improves slightly on the causal case. In the context of Opus-coded
speech, the proposed PLC significantly out-performs the existing Opus
PLC.

\section{Conclusions}

We have demonstrated that by splitting packet loss concealment into
a feature-wise prediction model and a sample-wise generative model,
we can achieve high-quality concealment that surpasses existing conventional
PLC algorithms. The proposed solution operates in real-time on 10\nobreakdash-ms
frames, with or without look-ahead, using less than 14\% of a laptop
CPU core. We have also demonstrated that our work can be applied in
the context of a modern speech compression algorithm and also out-perform
conventional concealment.

\balance

\bibliographystyle{IEEEtran}
\bibliography{plc,corpora}

\end{document}